\documentclass[12pt]{article}
\usepackage{amsmath}
\usepackage{amssymb}
\usepackage{amsfonts}
\usepackage{amscd}
\usepackage{delarray}
\usepackage{graphicx}

\usepackage{amsmath,amsmath,amsfonts,amssymb,color,bbm}

\usepackage{graphicx}

\def\Symp#1,#2,#3,#4.{\left[\left(\begin{array}{c}#1\\#2\end{array}\right),\left(\begin{array}{c}#3\\#4\end{array}\right)\right]}
\def\Vec#1,#2.{\left(\!\begin{array}{c}#1\\#2\end{array}\!\right)}
\def\vec#1,#2.{{#1\choose{#2}}}
\def\ket#1.{|#1\rangle}

\def\bra#1.{\langle#1|}
\def\braket#1,#2.{\langle#1|#2\rangle}
\newcommand{\beq}{\begin{equation}}
\newcommand{\eeq}{\end{equation}}
\newcommand{\beqa}{\begin{eqnarray}}
\newcommand{\eeqa}{\end{eqnarray}}

\begin{document}

\title{ Correlations of decay times of entangled composite unstable systems.}
\date{}
\author{}
\maketitle

\centerline{ Thomas Durt\footnote{Institut Fresnel, Domaine Universitaire de Saint-J\'er\^ome,\\ Avenue Escadrille Normandie-Ni\'emen, 13397 \\Marseille Cedex 20, France\\thomas.durt@centrale-marseille.fr}}

\begin{abstract}
The role played by Time in the quantum theory is still mysterious by many aspects. In particular it is not clear today whether the distribution of decay times of unstable particles could be described by a Time Operator. As we shall discuss, different approaches to this problem (one could say interpretations) can be found in the literature on the subject. As we shall show, it is possible to conceive crucial experiments aimed at distinguishing the different approaches, by measuring with accuracy the statistical distribution of decay times of entangled particles. Such experiments can be realized in principle with entangled kaon pairs.

\end{abstract}

\section{Introduction.}The role of Time in the quantum theory has not been fully clarified yet.\cite{booktime} According to many authors, time is an external parameter that cannot be put on the same footing as space. On the other side, relativity theory teaches us that beside the three pairs of conjugated quantitites $(x,p_x)$, $(y,p_y)$, and $(z,p_z)$, it is natural to consider a fourth pair, $(t,E)$. Nevertheless, such an idea causes problems because, as was noted by Pauli, it contradicts the fact that in nature the operators associated to energy (hamiltonians) are bounded by below\footnote{This is because time would be the generator of translations of energy of arbitrary-positive and/or negative-magnitudes.}. Due to Pauli's objections, many physicists doubt nowadays about the possibility that time is described by a self-adjoint operator conjugated to the hamiltonian, the so-called Time Operator.

To illustrate the problems encountered when we treat time as a measurable quantity, let us consider the so-called screen problem\cite{mielnik}, in which a quantum particle is prepared in a certain region of space (the left region) and propagates towards another region (the right region). To simplify the presentation, we shall consider that the particle propagates along a one dimensional axis ($X$) of which the origin lies in-between the left ($x<0$) and right ($x>0$) regions. If we know the wave function $\Psi(x,t)$ for all positions at time $t$, we are able, making use of the Born rule, to predict the probability to find, at time $t$, the particle in the interval $[x,x+dx]$, and we find that this probability is equal to $|\Psi(x,t)|^2\cdot dx$. In particular, the probability to find the particle in the left (right) region is equal to $\int_{-\infty}^0dx|\Psi(x,t)|^2$ ($\int^{+\infty}_0dx|\Psi(x,t)|^2$). What the quantum theory {\it per se\footnote{\label{bohm}If we consider the bohmian interpretation\cite{cushing}, which is a particular interpretation of the quantum theory, this is no longer true because in the de Broglie-Bohm approach particles are assumed to have a well-defined position and velocity at all times. In this approach the probability that the particle passes from the left to the right region between time $t$ and time $t+dt$ is equal to ${\hbar\over m}Im.(\Psi^*(x=0,t)\nabla_x\Psi(x=0,t))\cdot dt$. This quantity is properly normalised when the full wave packet entirely passes from the left region to the right region after a sufficiently long time (no echo and/or revival).}} does not allow us to estimate unambiguously however is the time at which the particle will pass from the left to the right region.

Similarly, if we consider a particle that is trapped between barrers of potential it is not clear how to estimate on the basis of the quantum theory at what time the particle will escape (tunnel) out of the region where it is confined by the potential. This is also the case if we consider unstable particles: there is no strict rule that allows us to derive the distribution of the times at which such particles will decay, although there exist different {\it recipes} that allow us in principle to do so. It is our goal in the present paper to describe three such recipes (section 2) that all lead to the same predictions in the case of exponential, single particle, decay. However, as has been discussed elsewhere\cite{09temporal,champ}, a fine structure appears in the case of non-exponential single particle decay (section 3). In this paper we pay particular attention to the case of two entangled unstable particles (section 4), and show that also in that case a fine structure appears regarding the predictions of the decay times distribution. In particular, entangled kaon pairs, that are nowadays produced in different places in the world\cite{frascati}, constitute good candidates for testing the kind of effects that we have in mind.
\section{Three approaches to Decay Time Distributions.}
\subsection{The standard approach.}
What we call the standard approach can be traced back to Fermi's derivation of his famous golden rule\cite{englert}, and to the celebrated Wigner-Weisskopf derivation\cite{WW}. Typically, those approaches deal with situations where a system, that is initially (at time 0) prepared in an unstable state, undergoes irreversible transitions to a continuum of final states. Their aim is to estimate the survival probability of the system at time $t$. In such cases, it is convenient to express the Hilbert space of the full system (unstable state plus decay products) as the direct sum of the Hilbert space of the ``surviving system'' and the Hilbert space of the decay products. For instance, when the system is a double level atom coupled to a continuum of electro-magnetic modes (that we assume to be prepared at temperature 0, that is, in the vacuum state, for simplicity), the initial state of the system can be described as follows:

$|\Psi(t=0)>=|e>^A\otimes |0>^L$ where $|e>^A$ represents the ``upper'' (excited) atomic state, while $|0>^L$ represents the unpopulated (vacuum) light state. When time flows, the amplitude of the initial state diminishes while the amplitudes of states of the type $|g>^A\otimes |1,{\bf k}>^L$ increase, where

$|g>^A$ represents the atomic ``lower'' (ground) state and $|1,{\bf k}>^L$ corresponds to the creation of 

one photon in a mode of wave vector ${\bf k}$ (energy conservation imposes that ${h\over \lambda}\cdot c= E_e-E_g$, with $c$ the speed of light and $E_e$ ($E_g$) the energy of the excited (ground) atomic state.

In summary, evolution sends the initial state $|\Psi(t=0)>=|e>^A\otimes |0>^L$ onto $|\Psi(t)>=\psi^{standard}(t)|e>^A\otimes |0>^L+\Sigma_{{\bf k}}\Psi_{{\bf k}}(t)|g>^A\otimes |1,{\bf k}>^L.$

In this context it is natural to interpret the survival probability $P_s(t)$ as $|\psi(t)|^2$. Standard treatments {\it \`a la} Fermi and/or {\it \`a la} Wigner-Weiskopff\cite{englert} show that in good approximation $|\psi(t)|^2$ decreases exponentially with time.

Formally, $P_s(t)=|<\Psi(t=0)|\Psi(t)>|^2$, so that the survival probability can be interpreted as the overlap between the initial state and the state at time $t$, but this is not always true. Let us consider for instance the case where initially the atom is prepared in a superposition of two excited states, so that $|\Psi(t=0)>=\psi_1(t=0)|e_1>^A\otimes |0>^L+\psi_2(t=0)|e_2>^A\otimes |0>^L$, and that these excited states decay into different ground states; then the state at time $t$ obeys 

\begin{eqnarray}\nonumber |\Psi(t)>=\psi_1^{standard}(t)|e_1>^A\otimes |0>^L+\psi^{standard}_2(t)|e_2>^A\otimes |0>^L+\\ \Sigma_{{\bf k}}\Psi_{{\bf k}}(t)|g_1>^A\otimes |1,{\bf k}>^L+\Sigma_{{\bf k'}}\Psi_{{\bf k'}}(t)|g_2>^A\otimes |1,{\bf k}>^L\end{eqnarray} 

and (provided the wave function is properly normalised at time $t=0$: $\Sigma_{j=1,2}| \psi_j(t=0)|^2$=1) the survival probability $P_s(t)$ is equal to $|\psi^{standard}_1(t)|^2$+$|\psi^{standard}_2(t)|^2$. 

It is no longer true that $P_s(t)=|<\Psi(t=0)|\Psi(t)>|^2$, but the survival probability can now be interpreted as the overlap between the Hilbert space assigned to the surviving components and the state at time $t$. Typically, in the Wigner-Weisskopf regime, $\psi^{standard}_j(t)=\psi^{standard}_j(t=0)exp(-i{ E_j t\over h})exp(-{\Gamma_j t\over 2 })$, where $\Gamma_j$ is the decay rate (that is the inverse of the life time) of the energy level $j$ ($j=1,2$).

Then the survival probability obeys 

\begin{equation}P_s(t)=\Sigma_{j=1,2}| \psi^{standard}_j(t=0)exp(-t({ i E_j\over h}+{\Gamma_j \over 2 }))|^2,\end{equation}

This behaviour also characterizes kaons, at least in first approximation, when we can neglect CP-violation effects\cite{cds2}. It is indeed common to treat kaons in the iso-spin picture, where their state is identical to the state of a two-level system\cite{perkins}. Each kaon state is then seen as a superposition of $|K_1>$ and $|K_2>$ states that are respectively eigenstates of the CP operator for the eigenvalues +1 and -1. It is also possible to discriminate  experimentally whether the decay products are  + or -1 CP-eigenstates and thus to evaluate experimentally the decay rate in function of time in each CP sector\footnote{Actually, $\mathrm{K}_1$
and $\mathrm{K}_2$ are in first approximation (in absence of
$CP$-violation) the decay modes of the kaons so that the weak disintegration process distinguishes the
$\mathrm{K}_{1}$ states which decay only into ``$2\pi$'' while the
$\mathrm{K}_{2}$ states decay into ``$3\pi, \pi e \nu, ...$''
\cite{perkins}. The lifetime of the $\mathrm{K}_{1}$ kaon is short
($\tau_{S}\approx 8.92\times10^{-11}~^\mathrm{s}$), while the
lifetime of the $\mathrm{K}_{2}$ kaon is quite longer
($\tau_{L}\approx 5.17\times10^{-8}~^\mathrm{s}$)\cite{perkins}.}. In particular, when kaons originate from a well-located source, the distance between the source and the place where detectors measure the decay products is proportional to the time elapsed since the creation of the kaon ($t=0$). Distance serves then as a clock, and it is possible to measure the distribution of decay times. In first approximation, we can neglect CP-violation, in which case the probability per unit of time or density of probability that a single kaon decays in the CP=+1 (-1) sector (from now on denoted $p_{d,1(2)}(t)$) is an exponential function of time:

\begin{eqnarray}\nonumber p^{standard}_{d,j}(t)=-{d| \psi^{standard}_j(t)|^2\over dt}=-{d| \psi^{standard}_j(t=0)exp(-t({ i E_j\over h}+{\Gamma_j \over 2 }))|^2\over dt}\\=| \psi^{standard}_j(t=0)|^2\Gamma_j exp(-\Gamma_j t), j=1,2\end{eqnarray}

\subsection{The hybrid approach.} In order to introduce properly the hybrid approach, let us reconsider the screen problem\cite{mielnik} that we addressed in the introduction, during which we aim at evaluating at which time a quantum particle, initially located in the left region ($x<0$) and moving in the direction of the right region ($x>0$) will pass through the origin. We found (cfr footnote \ref{bohm}) that in the bohmian approach, the corresponding temporal density of probability is equal to ${\hbar\over m}Im.(\Psi^*(x=0,t)\nabla_x\Psi(x=0,t)).$ Now, if we integrate this quantity from $x=-\infty$ to $x=+\infty$, we find after some simple computation that $\int_{-\infty}^{+\infty} dx {\hbar\over m}Im.(\Psi^*(x,t)\nabla_x\Psi(x,t))=\int_{-\infty}^{+\infty}dx \Psi^*(x,t){\hbar\nabla_x\over im}\Psi(x,t)=<v_x(t)>$. Considered so, ${\hbar\over m}Im.(\Psi^*(x,t)\nabla_x\Psi(x,t))$ can be seen as a local density of current, an interpretation that fits with the Madelung and de Broglie-Bohm interpretations\cite{cushing} according to which the local density of position is, at time $t$, $|\Psi(x,t)|^2$ and the local velocity is ${{\hbar\over m}Im.(\Psi^*(x,t)\nabla_x\Psi(x,t))\over |\Psi(x,t)|^2}$. The density of probability (per unit of time) that the particle passes from the left to the right region can thus be seen as a product of two quantities:

\begin{equation}p^{x=0}_d(t)={{\hbar\over m}Im.(\Psi^*(x,t)\nabla_x\Psi(x,t))\over |\Psi(x,t)|^2}\cdot |\Psi(x=0,t)|^2\label{troie}\end{equation}

These two quantities are represented in orthodox quantum mechanics by two non-commuting observables, velocity and position, so that they are in principle not measurable simultaneously.  This explains why in the Copenhagen interpretation it is not ``politically correct'' to assign well-defined trajectories to quantum particles. Even the product of these quantities, that is, the local density of probability per unit of time $p^{x=0}_d(t)$ 
is difficult to associate to a measurable quantity (observable) in the orthodox approach because it is not clear at all how to associate it to a self-adjoint operator. Nevertheless, in a heuristic approach, the quantity $p^{x=0}_d(t)\cdot dt$ can be related intuitively, in virtue of equation \ref{troie}, to the ``joint-probability'' that the particle arrives at the origin at time $t$ AND then passes from the left to the right during the time interval $dt$ (the first step is proportional to $|\Psi(x=0,t)|^2$ and the second step occurs with a frequency proportional to the ``local velocity'').

The ``hybrid'' approach is characterized by a similar intuition\cite{arxiv}: the probability that a kaon decays at time $t$ in the CP=+1 (-1) factor is predicted in this approach to be equal to the product of the ``survival probability projected in the corresponding CP sector'' (that can be seen in a semi-classical picture as the ``population'' of $|K_j>$ states, $j=1,2$) multiplied by the corresponding decay rate:

 \begin{equation}p^{hybrid}_{d,j}(t)=| \psi^{standard}_j(t)|^2\Gamma_j, j=1,2\end{equation}
In the case of exponential decay (that is when it is consistent to neglect CP-violation), the standard and hybrid approaches are de facto equivalent:

\begin{eqnarray}p^{standard}_{d,j}(t)\nonumber =-{d| \psi^{standard}_j(t=0)exp(-t({ i E_j\over h}+{\Gamma_j \over 2 }))|^2\over dt}\\=| \psi^{standard}_j(t=0)|^2\Gamma_j exp(-\Gamma_j t)=| \psi^{standard}_j(t)|^2\Gamma_j=p^{hybrid}_{d,j}(t), j=1,2\end{eqnarray}

It is worth noting that, most often implicitly, the majority of particles physicists resort, in the framework of meson phenomenology,  to the ``hybrid'' approach.\cite{christ,bertlmann,frascati,PRA63}
\subsection{Time Operator approach.}
The basic concept in the Time Operator (T.O.) approach is that one can, in analogy with what is done regarding position in the first quantization procedure, associate a wave function to the ``observable'' time of decay, on the basis of the substitutions

i) Position $x$ $\leftrightarrow$ Time of Decay $t$

ii) Spatial Wave Function $\psi(x)$  $\leftrightarrow$ Temporal Wave Function $\psi(t)$

iii) Probability to find the particle between $x$ and $x+dx$=  $|\psi(x)|^2dx$ $\leftrightarrow$ Probability that the particle decays between $t$ and $t+dt$=  $|\psi^{T.O.}(t)|^2dt$.

 This approach has been commented by us in more details in other papers\cite{09temporal,arxiv}. Despite of the fact that it suffers from Pauli's aforementioned objections  it is undirectly supported by the empirical evidence that the uncertainty on the energy of the initial state and the uncertainty on the time at which the decay occurs obey an Heisenberg-like complementarity relation.\cite{busch} This is observed for instance in spectroscopy where line-widths $\delta \nu$ of spectroscopical transitions and lifetimes $\delta \tau={1\over \Gamma}$ of the corresponding metastable states obey an uncertainty relation of the type $\delta \nu \delta \tau\approx 1.$ This is also observed in particle physics where the measure of the distribution of energies of the decay products of an unstable particle provides an alternative way to measure its $\Gamma$ factor.\cite{champ}

\section{Discriminating the three approaches in the case of single particle decay.} In the case of exponential decay, one cannot discriminate between the three approaches.\cite{09temporal,arxiv,champ} For instance if we choose 

$\psi^{T.O.}_j(t)=\psi^{standard}_j(t=0)\sqrt{\Gamma_j}exp(-t({ i E_j\over h}+{\Gamma_j \over 2 }))$ we find 
 \begin{equation}p^{T.O.}_{d,j}(t)=| \psi^{T.O.}_j(t)|^2=| \psi^{standard}_j(t=0)|^2\Gamma_j exp(-\Gamma_j t)=p^{standard}_{d,j}(t), j=1,2,\end{equation}in agreement with the standard and hybrid approaches.
 
 In the case of non-exponential decay however this is no longer true. In particular this is not true when quantum beats occur (these quantum beats can be observed in optics \cite{scully} but also in kaon phenomenology where CP-violation induces quantum beats in the temporal distributions of decay in the respective CP=+1 and -1 sectors \cite{perkins,christ}).
 
 It is not our goal to discuss this situation extensively in the present paper, because we addressed it already in the past. In summary, when two exponential decay processes coherently interfere, the intensity of the interference (beat) would make it possible to discriminate between the three approaches. For instance \cite{champ}, when we superpose two processes 1 and 2 characterized by complex amplitudes $\alpha_1, \alpha_2$, real energies $E_1, E_2$ and gamma factors $\Gamma_1, \Gamma_2$, we find that the ``standard'' decay rate $p^{standard}_d(t)$, defined by $-dP^{standard}_s(t)/dt$, obeys
\begin{equation}
p^{standard}_d(t)=\label{superstandard}N\cdot (| \alpha_1|^2\Gamma_1e^{-\Gamma_1t}+|\alpha_2|^2\Gamma_2e^{-\Gamma_2t}+2| \alpha_1||\alpha_2|e^{- \bar \Gamma t}R\cos(\Delta Et+\Delta \phi+\theta)),\end{equation}
where $R$ and $\theta$ are real and obey $Re^{i\theta}= \bar \Gamma-i\Delta E$, with $\bar \Gamma={\Gamma_1+\Gamma_2\over 2}$, $\Delta E=E_2-E_1$, $\Delta \phi= phase({\alpha_2\over \alpha_1})$, while $N$ is a normalisation factor (with $h=1$).

In the hybrid approach we find

\begin{equation} 
p^{hybrid}_d(t)=\label{hybrid}N^{'}(| \alpha_1|^2e^{-\Gamma_1t}+|\alpha_2|^2e^{-\Gamma_2t}+2| \alpha_1||\alpha_2|e^{- \bar  \Gamma t}\cos(\Delta Et+\Delta \phi)).\end{equation}
where $N'$ is a normalisation factor.

In the time operator approach we get

\begin{equation}
p^{TO}_d(t)= N^{''}\cdot (| \alpha_1|^2\Gamma_1e^{-\Gamma_1t}+|\alpha_2|^2\Gamma_2e^{-\Gamma_2t}+2| \alpha_1||\alpha_2|\sqrt{\Gamma_1\Gamma_2}e^{- \bar  \Gamma t}\cos(\Delta Et+\Delta \phi))
\label{timeoperator},\end{equation}where $N^{''}$ is a normalisation factor.
Obviously, the three expressions considered above generally differ from each other, which opens the door to experimental discriminations.\cite{09temporal,arxiv,champ} Now we shall examine to which extent this is also true in the case of entangled kaon pairs.
\section{Discriminating the three approaches in the case of entangled kaon pairs.}
\subsection{EPR kaon states}
EPR correlations\cite{EPR,Bell} exhibited by entangled mesons have been measured experimentally in several labs, (for instance in Frascati\cite{frascati} and in Geneva (CP-Lear\cite{cplear}) with kaons ($K$), and in Tsukuba (KEK\cite{tsukuba}) with $B$ mesons). 

 In the $\phi$ factory at Frascati for instance\cite{frascati}, the so-called $\phi$ resonance is produced. It has the property that during this process, kaon pairs are produced in the  EPR-Bohm (so-called {\it singlet}) state
 
 \begin{eqnarray}|\phi\rangle={1\over \sqrt 2}(|\mathrm{K}_1\rangle_l |\mathrm{K}_2\rangle_r-|\mathrm{K}_1\rangle_l|\mathrm{K}_2\rangle_r)\label{mix}  \nonumber\\
 ={1+|\epsilon|^2\over \sqrt 2}\bigg{(}| \mathrm{K}_S\rangle_l \mathrm{K}_L\rangle_r - | \mathrm{K}_L\rangle_l | \mathrm{K}_S\rangle_r\bigg{)},\label{phi}\end{eqnarray}
where the indices $l$ and $r$ refer to the fact that those kaons are sent along opposite directions (with equal velocities $v$). The factor $\epsilon$ reveals the presence of CP-violation effects. In practice, $\epsilon$ is a small parameter, and the ``real'' decay modes $| \mathrm{K}_S\rangle$ and $| \mathrm{K}_L\rangle$ are close to $|\mathrm{K}_1\rangle$ and $|\mathrm{K}_2\rangle$. 

Actually, the precise connection between the real modes ($S$ for short and $L$ for long) and the CP eigenstates (1 for CP=+1 and 2 for CP=-1) is the following:

\begin{equation}\label{CPviolS}
|K_S>= {1\over \sqrt{1+|\epsilon|^2}}(|K_1>+\epsilon |K_2>) 
\end{equation}

\begin{equation}\label{CPviolL}
|K_L>= {1\over \sqrt{1+|\epsilon|^2}}(|K_2>+\epsilon |K_1>),
\end{equation}

where $\epsilon$ the CP-violation parameter is a small parameter:

\begin{equation}\label{expepsilon}
|\epsilon|=(2.27\pm0.02)\times10^{-3}, ~~~
\mathrm{arg}(\epsilon)=43.37^\circ.
\end{equation}

The singlet state  state $|\phi\rangle$ possesses a total quasi-spin and a total momentum both equal to zero, due to conservation laws. Entanglement and EPR correlations are seen here to be the consequence of conservation laws, a common situation in quantum physics.

It has been shown recently\cite{catalina} that, despite of the fact that EPR correlations decay with time, it is nevertheless possible to violate well-chosen Bell inequalities with such states, so that their correlations do not admit a local realistic explanation. This work\cite{catalina} opens the way to a loophole-free violation of Bell's inequalities.

 In the next section we shall introduce a new proposal for generalizing the standard approach to the two particles case and show that in principle the correlations exhibited by the singlet state make it possible to discriminate the standard approach at one side, and the Time Operator and hybrid approach at the other side. Moreover, there exists an infinity of entangled states for which these results are still valid as we show in section\ref{nosinglet}.

\subsection{Standard and hybrid approach\label{manak}}
By a straightforward generalization of the previous treatment, we find that the standard prediction for the survival probability of the pair of kaons at distances $d_l=v\cdot t_l$ on the left and $d_r=v\cdot t_r$ can be expressed as follows, after expanding the wave function in the CP product eigen-basis:

\begin{equation}P^{standard}_S(t_l,t_r)=\|\psi(t_l,t_r)^2\|/\|\psi(0,0)^2\| \end{equation} with \begin{equation}\| \psi(t_l,t_r)\|^2=\sum_{i,j=1}^2\psi^*_{ij}(t_l,t_r)\psi_{ij}(t_l,t_r),\end{equation} where (taking $h=c=1)$ 

\begin{eqnarray}\psi_{ij}(t_l,t_r)=_l\langle\mathrm{K}_i| _r\langle\mathrm{K}_j|\psi(t_l,t_r)\rangle\nonumber\\=_l\langle\mathrm{K}_i| _r\langle\mathrm{K}_j|{1+|\epsilon|^2\over \sqrt 2}(e^{-\mathrm{i} (m_L-\frac{\mathrm{i}}{2}\Gamma_L)
t_l}|{\mathrm{K}}_L\rangle_l e^{-\mathrm{i} (m_S-\frac{\mathrm{i}}{2}\Gamma_S)
t_r}|\mathrm{K}_S\rangle_r-\nonumber\\e^{-\mathrm{i} (m_S-\frac{\mathrm{i}}{2}\Gamma_S)
t_l}|{\mathrm{K}}_S\rangle_l e^{-\mathrm{i} (m_L-\frac{\mathrm{i}}{2}\Gamma_L)
t_r}|\mathrm{K}_L\rangle_r)\nonumber\\
={1+|\epsilon|^2\over \sqrt 2}(e^{-\mathrm{i} (m_L-\frac{\mathrm{i}}{2}\Gamma_L)
t_l}e^{-\mathrm{i} (m_S-\frac{\mathrm{i}}{2}\Gamma_S)
t_r} \langle\mathrm{K}_i|{\mathrm{K}}_L\rangle_l  \langle\mathrm{K}_j|\mathrm{K}_S\rangle_r
\nonumber\\-e^{-\mathrm{i} (m_S-\frac{\mathrm{i}}{2}\Gamma_S)
t_l} e^{-\mathrm{i} (m_L-\frac{\mathrm{i}}{2}\Gamma_L)
t_r} \langle\mathrm{K}_i|{\mathrm{K}}_S\rangle_l  \langle\mathrm{K}_j|\mathrm{K}_L\rangle_r
\label{shishi}\end{eqnarray}

Interpolating from the previous analysis made in the single particle case, it is natural to interpret the probability $P_S(t_l,t_r)$ as the sum of the probabilities that the pair survives without decaying in the CP=+1 or - 1 channels during the time interval $[0,t_l]$ in the left region and $[0,t_r]$ in the right region. 

Accordingly, the probability that for instance the pair decays in the CP=+1 left and right channels during the time intervals   $[t_l,t_l+\delta t_l]$ in the left  region and $[t_r,t_r+\delta t_r]$ in the right region ought to be related to the variation of the projected survival probability  $P^{11}_S(t_l,t_r)$ during these periods of time. This variation is equal to 
\begin{equation}\delta P^{11}_S(t_l,t_r)\label{delat}=(\delta t_l\frac{\partial }{\partial t_l}+\delta t_r\frac{\partial }{\partial t_r})P^{11}_S(t_l,t_r)+\delta t_l\delta t_r \frac{\partial^2 }{\partial t_r\partial t_l}P^{11}_S(t_l,t_r),\end{equation}plus higher order terms in $\delta t_l$ and $\delta t_r$.

In a previous approach we neglected the quadratic contribution, and assumed that, for reasons of symmetry   $\delta t_l$=$\delta t_r$. We sketch in appendix the results that can be derived in this approach, but in the meanwhile, we met some doubts about its validity\cite{thanks}, which led us to introduce in the present paper a new definition of the standard density of probability of decay that we shall study in the coming sections.

Intuitively, we can motivate this new definition as follows: passing from the single particle case to the two particles case, one should find a genuine density of probability that is defined in the quadrant $[t_l>0,t_r>0]$. Formally, this density can be obtained by deriving the infinitesimal variation $\delta P^{11}_S(t_l,t_r)$ estimated in equation (\ref{delat}) relatively to the ``infinitesimal element of surface'' $\delta t_l \cdot \delta t_r$. By doing so we get

\begin{equation}p^{11standard}_d(t_l,t_r)={\delta P^{11}_S(t_l,t_r)\over \delta t_l\delta t_r}\label{delatnew}=\frac{\partial^2 }{\partial t_r\partial t_l}P^{11}_S(t_l,t_r),\end{equation}

This is a reformulation of the standard approach in the two particles case that is more natural in a sense than the previous one\cite{arxiv} (commented in appendix) because it implies densities by squared times and not by time. Moreover, it possesses the enjoyable property
 \begin{equation}\int_{t_l^1}^{t_l^2}d t_l\int_{t_r^1}^{t_r^2}dt_r p^{11standard}_d(t_l,t_r)=P^{11}_S(t_l^2,t_r^2)-P^{11}_S(t_l^1,t_r^1),\end{equation}in virtue of which it also has the advantage to be properly normalised: 

 \begin{equation}\int_0^{+\infty}d t_l\int_0^{+\infty}dt_r p^{11standard}_d(t_l,t_r)=P^{11}_S(t_l=0,t_r=0)\end{equation}

 A direct computation shows that, when the pair of kaons is prepared in the singlet state,

\begin{eqnarray}&p^{11standard}_d(t_l,t_r)&=(\frac{\partial }{\partial t_l}\frac{\partial }{\partial t_r})P^{11standard}_S(t_l,t_r)\label{frascatipd}\\&\approx& (\frac{\partial }{\partial t_l}\frac{\partial }{\partial t_r}){|\epsilon|^2\over  2} 
|e^{-\mathrm{i} (m_L-\frac{\mathrm{i}}{2}\Gamma_L)
t_l}e^{-\mathrm{i} (m_S-\frac{\mathrm{i}}{2}\Gamma_S)
t_r} -
e^{-\mathrm{i} (m_S-\frac{\mathrm{i}}{2}\Gamma_S)
t_l} e^{-\mathrm{i} (m_L-\frac{\mathrm{i}}{2}\Gamma_L)
t_r}|^2
\nonumber\\ &=& |\epsilon|^2  \cdot  \bigg{(}
\Gamma_S\cdot \Gamma_L\cdot(e^{-\Gamma_Lt_l-\Gamma_S t_r} +e^{-\Gamma_St_l-\Gamma_L t_r})
\nonumber \\&-&2e^{-\frac{(\Gamma_S+\Gamma_L)(t_l+t_r)}{2}}
[(\frac{(\Gamma_S+\Gamma_L)}{2})^2+(\Delta m)^2]cos (\Delta m (t_l-t_r))\bigg{)}
\nonumber\end{eqnarray}

In the case of kaons, $\frac{(\Gamma_S+\Gamma_L)}{2})\approx \Delta m\approx \frac{\Gamma_S}{2}$, so that
\begin{equation}p^{11standard}_d(t_l,t_r)\approx |\epsilon|^2 
\cdot  \bigg{(}
\Gamma_S\cdot \Gamma_L\cdot(e^{-\Gamma_Lt_l-\Gamma_S t_r} +e^{-\Gamma_St_l-\Gamma_L t_r})
\nonumber \\-\Gamma_S^2e^{-\frac{(\Gamma_S+\Gamma_L)(t_l+t_r)}{2}}
cos (\Delta m (t_l-t_r))\bigg{)}\end{equation}

In the two particles case, the straightforward generalisation of the hybrid approach leads us to define  \begin{equation}p^{ijhybrid}_d(t_l,t_r)=\Gamma_i\Gamma_jP^{ij}_S(t_l,t_r)\label{delathybrid}\end{equation} so that

\begin{eqnarray}p^{11hybrid}_d(t_l,t_r)\approx \Gamma_S\Gamma_L|\epsilon|^2 |e^{-\mathrm{i} (m_L-\frac{\mathrm{i}}{2}\Gamma_L)t_l}e^{-\mathrm{i} (m_S-\frac{\mathrm{i}}{2}\Gamma_S)t_r} -e^{-\mathrm{i} (m_S-\frac{\mathrm{i}}{2}\Gamma_S)t_l} e^{-\mathrm{i} (m_L-\frac{\mathrm{i}}{2}\Gamma_L)
t_r}|^2\nonumber \\=\Gamma_S\Gamma_L|\epsilon|^2\bigg{(}
e^{-\Gamma_Lt_l-\Gamma_S t_r} +e^{-\Gamma_St_l-\Gamma_L t_r}
-2e^{-\frac{(\Gamma_S+\Gamma_L)(t_l+t_r)}{2}}
cos (\Delta m (t_l-t_r))\bigg{)}\nonumber\end{eqnarray}

It is worth noting that the former quantity is proportional (up to a calibration factor $|\langle \pi^+\pi^- | \mathrm{K}_1\rangle|^4  |$ aimed at converting the kaonic decay rate in the CP=+1 sector into the detection rate of pion pairs)
to what is called in Ref.\cite{frascati} (expression (2)) ``{\it the decay intensity for the process}  $\phi\rightarrow ({\rm 2 neutral kaons})\rightarrow  \pi^+\pi^-$''.

Obviously, despite of the fact that the purely exponential contributions are the same in both approaches,

\begin{equation}p^{11standard}_d(t_l,t_r)\not=p^{11hybrid}_d(t_l,t_r)\end{equation}
 As in the single particle case\cite{champ}(equations (\ref{superstandard},\ref{hybrid})), interferences make it thus possible to distinguish the two approaches.

\subsection{Time Operator approach.\label{singletTO}}

Let us now consider the prediction that we get in the Time Operator approach. 

In this approach, when at time 0 a $|\mathrm{K}_S\rangle$ is prepared, it is described in the CP eigen basis $|K_1>=(1,0); |K_2>=(0,1)$, by the quasi-spinorial temporal wave function
\begin{equation}\label{psitS}
(\psi^S_1(t),\psi^S_2(t))= {\sqrt{1\over 1+|\epsilon|^2}}(1,\epsilon) \sqrt{\Gamma_S}e^{-\mathrm{i} (m_S-\frac{\mathrm{i}}{2}\Gamma_S)
t},
\end{equation}
based on the equality (\ref{CPviolS}). Similarly, in virtue of  (\ref{CPviolL}), a $\mathrm{K}_L$ state is associated to the bi-spinorial function ($\psi^L_1(t)$,$ \psi^L_2(t)$) through

\begin{equation}\label{psitL}
(\psi^L_1(t),\psi^L_2(t))= {\sqrt{1\over 1+|\epsilon|^2}}(\epsilon,1) \sqrt{\Gamma_L}e^{-\mathrm{i} (m_L-\frac{\mathrm{i}}{2}\Gamma_L)
t}
\end{equation}
Combining (\ref{psitS},\ref{psitL}) with (\ref{phi}) we get

 \begin{eqnarray}|\psi\rangle^{T.O.}(t_l,t_r)&\approx&
 {1\over \sqrt 2}\bigg{(}\left(\begin{array}{c} 1\\ \epsilon
\end{array}\right)_l\otimes 
\left(\begin{array}{c} \epsilon \\ 1
\end{array}\right)_r \sqrt{\Gamma_S}e^{-\mathrm{i} (m_S-\frac{\mathrm{i}}{2}\Gamma_S)
t_l}\cdot  \sqrt{\Gamma_L}e^{-\mathrm{i} (m_L-\frac{\mathrm{i}}{2}\Gamma_L)
t_r}-\nonumber \\
&&\left(\begin{array}{c} \epsilon \\ 1
\end{array}\right)_l
\otimes\left(\begin{array}{c} 1\\ \epsilon
\end{array}\right)_r \sqrt{\Gamma_L}e^{-\mathrm{i} (m_L-\frac{\mathrm{i}}{2}\Gamma_L)
t_l}\cdot  \sqrt{\Gamma_S}e^{-\mathrm{i} (m_S-\frac{\mathrm{i}}{2}\Gamma_S)
t_r}
\bigg{)}
\label{phinew}\end{eqnarray}

Projecting (\ref{phinew}) onto $\left(\begin{array}{c} 1\\ 0
\end{array}\right)_l\otimes 
\left(\begin{array}{c} 1 \\ 0
\end{array}\right)_r $ to evaluate the amplitude $\psi_{11}$ associated to the decay intensity for the process $\phi\rightarrow ({\rm 2 neutral kaons})\rightarrow  \pi^+\pi^-$ we get that, up to a global normalisation factor,

\begin{equation}\psi^{T.O.}_{11}\approx{1\over \sqrt 2}\epsilon\sqrt{\Gamma_S\Gamma_L}\bigg{(}e^{-\mathrm{i} (m_S-\frac{\mathrm{i}}{2}\Gamma_S)
t_l}\cdot  e^{-\mathrm{i} (m_L-\frac{\mathrm{i}}{2}\Gamma_L)t_l}-
e^{-\mathrm{i} (m_L-\frac{\mathrm{i}}{2}\Gamma_L)
t_l}\cdot e^{-\mathrm{i} (m_S-\frac{\mathrm{i}}{2}\Gamma_S)
t_r}
\bigg{)}
\end{equation}

Generalising the treatment of the single particle case, we impose that, in the Time Operator approach,  

\begin{equation}p^{ijT.O.}_d(t_l,t_r)=|\psi^{T.O.}_{ij}|^2\label{delatTO}\end{equation}so that the joint-probability of firing of left and right detectors in the CP=+1 sector is equal to the modulus squared of $\psi^{T.O.}_{11}$. Now, this quantity is proportional, up to a constant in time factor, to the quantity $P^{11}_S(t_l,t_r)$. 

\begin{equation}p^{11T.O.}_d(t_l,t_r)=N\cdot P^{11}_S(t_l,t_r)\end{equation}

Besides, \begin{equation}p^{11hybrid}_d(t_l,t_r)=N'\cdot P^{11}_S(t_l,t_r)\end{equation}

Normalisation\footnote{In last resort, normalisation is imposed by the calibration of the detectors. \label{cali}} imposes that $N=N'$ so that

\begin{equation}p^{11T.O.}_d(t_l,t_r)=p^{11hybrid}_d(t_l,t_r).\end{equation}

In conclusion, our analysis shows that by measuring the joint-probability of CP=+1 decay products, when the pair of kaons is prepared in the singlet state, it is not possible to discriminate the hybrid and Time Operator approaches, although their predictions differ from those made in the standard approach, a result that we generalize in the next sections (\ref{nosinglet},\ref{joint}).

 In the section \ref{nosingletno}, however, we shall show that it is not so for all entangled states, which opens the way to a new class of crucial experiments aimed at discriminating hybrid and Time Operator approaches.

\subsection{Generalisation to other entangled states ($\alpha$ states).\label{nosinglet}}

Let us now assume that kaon pairs are produced in the  EPR-Bohm $\psi^{\alpha}$ state defined by
 
 \begin{eqnarray}|\psi^{\alpha}\rangle
 ={1+|\epsilon|^2\over \sqrt 2}\bigg{(}| \mathrm{K}_L\rangle_l \mathrm{K}_S\rangle_r - e^{i\alpha}| \mathrm{K}_S\rangle_l | \mathrm{K}_L\rangle_r\bigg{)},\label{phialpha}\end{eqnarray}

then, following the same of reasoning as in the previous section we conclude that the projection on the CP=+1 channels of the survival probability obeys

\begin{eqnarray} P^{11\alpha}_S(t_l,t_r)\approx{|\epsilon|^2\over  2} 
|e^{-\mathrm{i} (m_L-\frac{\mathrm{i}}{2}\Gamma_L)
t_l}e^{-\mathrm{i} (m_S-\frac{\mathrm{i}}{2}\Gamma_S)
t_r} -
e^{i\alpha}e^{-\mathrm{i} (m_S-\frac{\mathrm{i}}{2}\Gamma_S)
t_l} e^{-\mathrm{i} (m_L-\frac{\mathrm{i}}{2}\Gamma_L)
t_r}|^2
\nonumber\\ = |\epsilon|^2 \cdot\bigg{(}
e^{-\Gamma_Lt_l-\Gamma_S t_r} +e^{-\Gamma_St_l-\Gamma_L t_r}-2e^{-\frac{(\Gamma_S+\Gamma_L)(t_l+t_r)}{2}}
cos (\Delta m (t_l-t_r)+\alpha)\bigg{)}\nonumber
\end{eqnarray}

By repeating the reasoning of the previous section we find that when kaon pairs are produced in the  EPR-Bohm $\psi^{\alpha}$ state, it is not possible to discriminate the hybrid and Time Operator approaches, although their predictions differ from those made in the standard approach.
\subsection{Other joint-detections.\label{joint}}So far, we only considered joint-detections in the left and right regions in the same CP sector (CP=+1). It is easy to convince oneself that our analysis remains unchanged when transposed to the (joint) CP=-1 detections.

If instead we consider, say, a CP=+1 detection at the right side and a CP=-1 detection at the left side, we get that, at the lowest order in $\epsilon$ (which is 0),  

$\psi_{12}(t_l,t_r)$ is proportional to $e^{-\mathrm{i} (m_L-\frac{\mathrm{i}}{2}\Gamma_L)
t_l}e^{-\mathrm{i} (m_S-\frac{\mathrm{i}}{2}\Gamma_S)
t_r}$. 

Then,

\begin{eqnarray} p^{12\alpha-standard}_d(t_l,t_r)=(\frac{\partial }{\partial t_l}\frac{\partial }{\partial t_r}) P^{12\alpha}_S(t_l,t_r)=\nonumber\\ (\frac{\partial }{\partial t_l}\frac{\partial }{\partial t_r})|e^{-\mathrm{i} (m_L-\frac{\mathrm{i}}{2}\Gamma_L)
t_l}e^{-\mathrm{i} (m_S-\frac{\mathrm{i}}{2}\Gamma_S)
t_r}|^2
\nonumber\\ =\Gamma_S \Gamma_LP^{12\beta}_S(t_l,t_r)=p^{12\alpha-hybrid}_d(t_l,t_r)\nonumber\\ =|\sqrt \Gamma_Le^{-\mathrm{i} (m_L-\frac{\mathrm{i}}{2}\Gamma_L)
t_l}\sqrt \Gamma_Se^{-\mathrm{i} (m_S-\frac{\mathrm{i}}{2}\Gamma_S)
t_r}|^2=p^{12\alpha-T.O.}_d(t_l,t_r)
\end{eqnarray}

A similar result holds when we permute the left and right region (or equivalently the +1 and -1 CP sectors). Henceforth, it is straightforward to check that for all for the $\psi^{\alpha}$ states (defined by equation (\ref{phi})), the hybrid, standard and Time Operator approaches cannot be discriminated through a measure of the statistical distribution of CP=+1 and -1 decay products in the left and right regions, at the order 0 in the parameter $\epsilon$, although the standard approach can be discriminated from the two other approaches in this case, by measuring correlations quadratic in $|\epsilon|$ as we have shown in a previous section (\ref{manak}). 

\subsection{$\beta$ states-other Bell states\label{nosingletno}}It is not true that for all entangled states one cannot discriminate between the hybrid and T.O. approaches. For instance, let us consider the state 
 \begin{eqnarray}|\psi^{\beta}\rangle
 ={1+|\epsilon|^2\over \sqrt 2(1-\epsilon^2)}\bigg{(}| \mathrm{K}_L\rangle_l \mathrm{K}_L\rangle_r - e^{i\beta}| \mathrm{K}_S\rangle_l | \mathrm{K}_S\rangle_r\bigg{)}\label{beta}\end{eqnarray}

These states are such that the standard joint-probability $ p^{11\beta}_d(t_l,t_r)$ of decay in the CP=+1 channels in the right and left regions obeys, at the dominating order in $\epsilon$ (which is now 0),

\begin{eqnarray} p^{11\beta-standard}_d(t_l,t_r)=(\frac{\partial }{\partial t_l}\frac{\partial }{\partial t_r}) P^{11\beta}_S(t_l,t_r)=\nonumber\\ (\frac{\partial }{\partial t_l}\frac{\partial }{\partial t_r})|-
e^{i\beta}e^{-\mathrm{i} (m_S-\frac{\mathrm{i}}{2}\Gamma_S)
(t_l+t_r)}|^2
\nonumber\\ =\Gamma_S^2P^{11\beta}_S(t_l,t_r)=p^{11\beta-hybrid}_d(t_l,t_r)=p^{11\beta-T.O.}_d(t_l,t_r)\end{eqnarray}

Similar results can be derived for the joint detections in the CP=-1 sector ($p^{22\beta}_d(t_l,t_r)$), showing that, at the order 0 in $\epsilon$, the three approaches lead to the same predictions.

Now, in this case, at the dominating order in $\epsilon$, the (hybrid) joint-probability $ p^{12\beta}_d(t_l,t_r)$ of decay in the CP=+1 channel in the right region and in the CP=-1 channel in the left region obeys

\begin{eqnarray} p^{12\beta-hybrid}_d(t_l,t_r)=N|\epsilon|^2|e^{-\mathrm{i} (m_L-\frac{\mathrm{i}}{2}\Gamma_L)
(t_l+
t_r)} -
e^{i\beta}e^{-\mathrm{i} (m_S-\frac{\mathrm{i}}{2}\Gamma_S)
(t_l+t_r)}|^2\nonumber \\=N|\epsilon|^2 \cdot\bigg{(}
e^{-\Gamma_L(t_l+t_r)} +e^{-\Gamma_S(t_l+t_r)}-2e^{-\frac{(\Gamma_S+\Gamma_L)(t_l+t_r)}{2}}
cos (\Delta m (t_l+t_r)+\beta)\bigg{)}\nonumber\end{eqnarray}

Besides, in the Time Operator approach, we find that  the temporal density of probability is proportional to $|\epsilon|^2|\Gamma_Le^{-\mathrm{i} (m_L-\frac{\mathrm{i}}{2}\Gamma_L)
(t_l+
t_r)} -\Gamma_S
e^{i\beta}e^{-\mathrm{i} (m_S-\frac{\mathrm{i}}{2}\Gamma_S)
(t_l+t_r)}|^2$, which is not proportional to the hybrid density so that these two approaches are likely to be discriminated in the present situation...

Similar results hold concerning $ p^{21\beta}_d(t_l,t_r)$.

This second class of maximally entangled states for which hybrid and Time Operator predictions obviously differ comprises the two remaining Bell states\cite{bellstate}

 ${1+|\epsilon|^2\over \sqrt 2(1-\epsilon^2)}\bigg{(}| \mathrm{K}_L\rangle_l \mathrm{K}_L\rangle_r +| \mathrm{K}_S\rangle_l | \mathrm{K}_S\rangle_r\bigg{)}$ and ${1+|\epsilon|^2\over \sqrt 2(1-\epsilon^2)}\bigg{(}| \mathrm{K}_L\rangle_l \mathrm{K}_L\rangle_r -| \mathrm{K}_S\rangle_l | \mathrm{K}_S\rangle_r\bigg{)}$.

\section{Conclusion, open questions.}

In this paper we considered two infinite families of entangled bipartite kaonic states, the $\alpha$ and $\beta$ states. Each of them comprises two so-called {\it Bell} states\cite{bellstate}.  We showed that the hybrid (\ref{delathybrid}) and Time Operator (\ref{delatTO}) approaches lead to the same predictions but differ in their predictions from the standard approach (\ref{delatnew}) when the entangled state belongs to the $\alpha$ class that comprises the {\it singlet} state (that has been realized experimentally in the past). On the contrary, the hybrid and Time Operator predictions differ from each other when the second entanglement-class is considered, which shows that in principle crucial experiments could help to discriminate between the different approaches.

We are aware however that the second class of states is hard and maybe impossible to prepare in a lab. today.\cite{catalina} Moreover, even in the class that comprises the singlet state, the effects that are involved are due to CP-violation and therefore they are weak effects. It is interesting because of this to check whether other strategies are worth of investigation, aiming at discriminating experimentally the three aforementioned approaches.

A first interesting alternative strategy is to investigate the possibility to perform active measurements.

Indeed, according to the distinction between active and passive measurements made in Ref.\cite{bramon} (see also our discussion of appendix 3 of Ref.\cite{arxiv}, and Ref.\cite{bramon2} for an application to Bell's inequalities), the Frascati experiment, in which the rates of production of pairs of pions are measured ALONG (and not ACROSS) the trajectories of the kaons in the left and right regions is a PASSIVE measurement. Now, nothing forbids to perform ACTIVE measurements, for instance, to interpose a slab of matter across the trajectories of the kaons in order to gain information about the $K_0$ and $\overline K_0$ populations. It is out of the scope of the present paper to tackle this problem, but obviously this strategy opens new perspectives regarding the discrimination between the different models presented in our paper.

It could well be after all that all the models presented here are wrong. For instance, we reproduce in appendix an alternative proposal, made in Ref.\cite{arxiv}, for the standard distribution (\ref{delatold},\ref{frascatipd}). It is easy to convince oneself that this approach does not bring anything new in absence of CP violation, compared to our new definition (\ref{delatnew}), but, who knows?, it could provide a better fit to experimental data than the three approaches studied previously whenever CP violation effects are present...

The Time Super Operator (TSO) approach\cite{cour80,MPC,09superop,TSO}  provides yet another challenging approach to the problem. It has already been shown elsewhere \cite{09superop,TSO} that the TSO predictions could also be discriminated from the standard, hybrid and T.O. approaches in the framework of kaon phenomenology, by measuring CP-violation effects at the single particle level, but a TSO treatment of the temporal statistics of decay of entangled kaon pairs has not been investigated yet.

It would also be worth considering quantum optical systems (unstable metastable states prepared in ion traps for instance\cite{champ}), where the experimental versatility (in the choice of the measurement basis and in the preparation of the initial state) and accuracy is quite higher than in particle physics. In physics, the last word ought to come to the experimentalists, always.

Last but not least, it would be interesting to consider the recent proposal for a violation of Bell inequalities\cite{catalina} having in mind the distinction between the three aforementioned approaches. It is worth noting, by the way, that in absence of CP violation, in the case of passive measurements, one is unable to discriminate between the three approaches (\ref{delatnew},\ref{delathybrid},\ref{delatTO}) and even the fourth one described in appendix (\ref{delatold},\ref{frascatipd}). This result is, maybe, linked to an older result according to which the violation of a certain class of Bell's inequalities is proportional to the magnitude of CP-violation\cite{PRA63}...but this is another story.

\section*{Acknowledgements}
Sincere thanks to R.A. Bertlmann, M. Courbage, C. Curceanu, B. Hiesmayr and M. Saberi for discussions on related subjects. Support from the COST action MP 1006 and the groupe Clart\'e (Institut Fresnel) is acknowledged.

\section{Appendix}
\subsection{``Old'' Standard approach}
In the ``old'' standard treatment of the two particles case, the probability that for instance the pair decays in the CP=+1 left and right channels during the time intervals   $[t_l,t_l+\delta t_l]$ in the left  region and $[t_r,t_r+\delta t_r]$ in the right region ought was related to the variation{\it -at the first order in $\delta t_l$ and $\delta t_r$-}of the projected survival probability  $P^{11}_S(t_l,t_r)$ during these periods of time. This variation is then to equal to 
\begin{equation}\delta P^{11}_S(t_l,t_r)\label{delatold}=(\delta t_l\frac{\partial }{\partial t_l}+\delta t_r\frac{\partial }{\partial t_r})P^{11}_S(t_l,t_r),\end{equation}plus higher order terms in $\delta t_l$ and $\delta t_r$.

If we neglect the quadratic contribution, and assume that, for reasons of symmetry\footnote{Assuming that detectors have the same dimensions at both sides, and also taking account of the fact that particles at both sides have same velocities so that $\tau_l=\tau_r$}, $\delta t_l$=$\delta t_r$, we find now that
\begin{equation}p^{11standard(old)}_d(t_l,t_r)=-(\frac{\partial }{\partial t_l}+\frac{\partial }{\partial t_r})P^{11}_S(t_l,t_r).\end{equation}

 In the case that the pair of Kaons is prepared in a {\it singlet} EPR-Bohm state, direct computation shows that 

\begin{eqnarray}&p^{11standard(old)}_d(t_l,t_r)&=-(\frac{\partial }{\partial t_l}+\frac{\partial }{\partial t_r})P^{11standard}_S(t_l,t_r)\label{frascatipd}\\&\approx& -(\frac{\partial }{\partial t_l}+\frac{\partial }{\partial t_r}){|\epsilon|^2\over  2} 
|e^{-\mathrm{i} (m_L-\frac{\mathrm{i}}{2}\Gamma_L)
t_l}e^{-\mathrm{i} (m_S-\frac{\mathrm{i}}{2}\Gamma_S)
t_r} -
e^{-\mathrm{i} (m_S-\frac{\mathrm{i}}{2}\Gamma_S)
t_l} e^{-\mathrm{i} (m_L-\frac{\mathrm{i}}{2}\Gamma_L)
t_r}|^2
\nonumber\\ &=& (\Gamma_S+\Gamma_L)|\epsilon|^2  \cdot\bigg{(}
e^{-\Gamma_Lt_l-\Gamma_S t_r} +e^{-\Gamma_St_l-\Gamma_L t_r}
-2e^{-\frac{(\Gamma_S+\Gamma_L)(t_l+t_r)}{2}}
cos (\Delta m (t_l-t_r))\bigg{)}
\nonumber\\ &=&(\Gamma_S+\Gamma_L){|\epsilon|^2\over  2} |e^{-\mathrm{i} (m_L-\frac{\mathrm{i}}{2}\Gamma_L)
t_l}e^{-\mathrm{i} (m_S-\frac{\mathrm{i}}{2}\Gamma_S)
t_r} -
e^{-\mathrm{i} (m_S-\frac{\mathrm{i}}{2}\Gamma_S)
t_l} e^{-\mathrm{i} (m_L-\frac{\mathrm{i}}{2}\Gamma_L)
t_r}|^2.\nonumber\end{eqnarray}

Remarkably, in this case, ${(\frac{\partial }{\partial t_l}+\frac{\partial }{\partial t_r})P^{11}_S(t_l,t_r)\over P^{11}_S(t_l,t_r)}$ is a constant factor that does not depend on time, so that the ``old'' standard and hybrid approach do effectively agree in this case. 

Putting this result together with the result of section \ref{singletTO} according to which the predictions made in the hybrid approach and in the T.O. approach are the same, we find that the ``old'' standard approach the hybrid and T.O. approach cannot be discriminated in the case of the singlet state. This conclusion is also valid for all $\alpha$ states (defined in section \ref{nosinglet}), a straightforward result that we mention without proof.

Now, if we consider the $\beta$ states defined in section \ref{nosingletno} (equation (\ref{beta})), at the dominating order in $\epsilon$, the ``old'' standard joint-probability $ p^{12\beta-standard(old)}_d(t_l,t_r)$ of decay in the CP=+1 channel in the right region and in the CP=-1 channel in the left region obeys

\begin{eqnarray} p^{12\beta-standard(old)}_d(t_l,t_r)&=&-(\frac{\partial }{\partial t_l}+\frac{\partial }{\partial t_r}) P^{12\beta}_S(t_l,t_r)\nonumber\\ &\approx&-(\frac{\partial }{\partial t_l}+\frac{\partial }{\partial t_r}){|\epsilon|^2\over  2} 
|e^{-\mathrm{i} (m_L-\frac{\mathrm{i}}{2}\Gamma_L)
(t_l+
t_r)} -
e^{i\beta}e^{-\mathrm{i} (m_S-\frac{\mathrm{i}}{2}\Gamma_S)
(t_l+t_r)}|^2
\nonumber\\ &=& {|\epsilon|^2\over 2} \cdot\bigg{(}
2\Gamma_Le^{-\Gamma_L(t_l+t_r)} +2\Gamma_Se^{-\Gamma_S(t_l+t_r)}\nonumber\\
&+&2e^{-\frac{(\Gamma_S+\Gamma_L)(t_l+t_r)}{2}}
[(\Gamma_S+\Gamma_L)cos (\Delta m (t_l+t_r)+\beta)
-\Delta m sin  (\Delta m (t_l+t_r)+\beta) ]    \bigg{)}
\label{frascatibeta}\nonumber \end{eqnarray}

This quantity is obviously NOT proportional to 

\begin{eqnarray}|\epsilon|^2|e^{-\mathrm{i} (m_L-\frac{\mathrm{i}}{2}\Gamma_L)
(t_l+
t_r)} -
e^{i\beta}e^{-\mathrm{i} (m_S-\frac{\mathrm{i}}{2}\Gamma_S)
(t_l+t_r)}|^2\nonumber \\=|\epsilon|^2 \cdot\bigg{(}
e^{-\Gamma_L(t_l+t_r)} +e^{-\Gamma_S(t_l+t_r)}-2e^{-\frac{(\Gamma_S+\Gamma_L)(t_l+t_r)}{2}}
cos (\Delta m (t_l+t_r)+\beta)\bigg{)}\nonumber\end{eqnarray}
Therefore the ``old'' standard and hybrid approaches are likely to be discriminated in the present situation.

Moreover, in the Time Operator approach, we find that  the temporal density of probability is proportional to $|\epsilon|^2|\Gamma_Le^{-\mathrm{i} (m_L-\frac{\mathrm{i}}{2}\Gamma_L)
(t_l+
t_r)} -\Gamma_S
e^{i\beta}e^{-\mathrm{i} (m_S-\frac{\mathrm{i}}{2}\Gamma_S)
(t_l+t_r)}|^2$, which is neither proportional to the hybrid density nor to the standard density so that the three approaches are likely to be discriminated in the present situation...

Similar results hold concerning $ p^{21\beta-standard(old)}_d(t_l,t_r)$.

Last but not least, it is worth mentioning another result that we do not prove explicitly but is easy to demonstrate: the ``old'' standard approach (\ref{delatold},\ref{frascatipd}) leads to exactly the same predictions as the ``new'' standard (\ref{delatnew}), hybrid (\ref{delathybrid}) and T.O. (\ref{delatTO}) approaches as far as we can neglect CP-violation effects, that is, at zero order in $|\epsilon|$.

\end{document}